# Galilean relativity, special relativity and the work-kinetic energy theorem revisited.

Bernhard Rothenstein, "Politehnica" University of Timisoara, Physics Department, Timisoara, Romania

*Abstract. Transformation equations for the kinetic energy of the same bullet in Galileo's and Einstein's special relativity are derived. The special relativity result is extended to the case of a photon.*

The Physics Teacher presents an interesting example of how Galilean relativity is involved in the presentation of a dynamics problem.[1] We present a simpler version of it using usual relativistic notations. Restating the problem consider two inertial reference frames I and I', I' moving with speed V relative to I in the positive direction of the overlapped OX(O'X') axes. The problem is to find out a relationship between the kinetic energies of the same ball moving along the same direction, as measured by observers from I and I' respectively. Making the convention to present physical quantities measured in I as unprimed and physical quantities measured in I' as primed, the work-kinetic energy theorem reads in I

$$dW = Fdx = Fudt \qquad (1)$$

reading in I'

$$dW' = F'dx' = F'u'dt' \qquad (2)$$

in accordance with the fact that the laws of mechanics are the same in all inertial reference frames in relative motion. In (1) and (2) W(W'), F(F'), u(u') and t(t') stand for the kinetic energy, force, speed and time respectively. Combining (1) and (2) we obtain

$$dW = dW' \frac{F}{F'} \frac{m}{m'} \frac{dt}{dt'}. \qquad (3)$$

In the limits of Newtonian mechanics, force, mass and time intervals have the same magnitude, the velocities adding as

$$u = u' + V \qquad (4)$$

the accelerations of the same particle having the same magnitude in both inertial reference frames, i.e.

$$a = a' = \frac{du}{dt} = \frac{du'}{dt'}. \qquad (5)$$

Taking all that into account (3) becomes successively

$$dW = dW'\left(1 + \frac{V}{u'}\right) = dW' + mVdu. \qquad (6)$$

Relativists would criticize (6) telling that in its right side we add a physical quantity measured in I'(W') with a physical quantity measured in I (mVdu) and so they present it as

$$dW = dW' + Vmdu'. \qquad (7)$$

Applied in the case of the scenario proposed followed in[1] (6) leads to

$$W = W' + mV(u_f - u_i) \qquad (8)$$

$u_f$ and $u_i$ standing for the final and the initial speeds of the ball in I recovering the results obtained in[1].

The authors of the paper we revise suggest that a relativistic version of the same problem could be of interest. In order to so solve the problem we start with (1) and (2) taking into account that the OX(O'X') component of the force has the same magnitude in all inertial reference frames[2] and that u and u' add as[2]

$$u = \frac{u' + V}{1 + \frac{u'V}{c^2}} \qquad (9)$$

and that time intervals transform as[2]

$$dt = \frac{dt' + \frac{V}{c^2} dx'}{\sqrt{1 - \frac{V^2}{c^2}}} \qquad (10)$$

the kinetic energies of the same ball transforming as

$$dW = dW' \frac{1 + \frac{V}{u'}}{\sqrt{1 - \frac{V^2}{c^2}}} \qquad (11)$$

with u'>0 in order to avoid indeterminacy. For u=u'=c the ball we considered so far is a photon the kinetic energy of which transforms as

$$dW_c = dW'_c \sqrt{\frac{1 + \frac{V}{c}}{1 - \frac{V}{c}}} \qquad (12)$$

We consider that our presentation is straightforward and time saving being consequently based on the concepts of classical and special relativity.

**References**

[1] Brandon J. Tefft and James A. Tefft, "Galilean relativity and the work-kinetic energy theorem," Phys.Teach **45,** 218-220 (2007)

[2] Robert Resnick, *Introduction to Special Relativity* (John Wiley & Sons, New York, London, Sydney 1968) pp. 56, 79 and 143.